\documentclass{article}
\usepackage{epsfig}
\usepackage{cite}

\usepackage{amssymb}
\usepackage{amsmath}
\usepackage{verbatim}
\usepackage{mathrsfs}
\usepackage{amsfonts}
\usepackage{latexsym}
\usepackage{epsfig}
\usepackage{color}
\usepackage{graphicx,subfigure}
\usepackage{units}
\usepackage{slashbox}
\usepackage{envmath}
\usepackage{bm,url}

\date{\today}

\usepackage{amsmath}
\usepackage{amssymb}

\definecolor{orange}{rgb}{0.9,0.45,0}

\long\def\symbolfootnote[#1]#2{\begingroup%
\def\thefootnote{\fnsymbol{footnote}}\footnote[#1]{#2}\endgroup}

\begin{document}


\title{\bf Self-interactions can stabilize excited boson stars}

\author{
{\large Nicolas Sanchis-Gual}$^{1}$,
{\large Carlos Herdeiro}$^{1}$, 
and
{\large Eugen Radu}$^{1}$
\\ 
$^{1}${\small Departamento  de  Matem\'{a}tica  da  Universidade  de  Aveiro }\\
{\small  and  Centre  for  Research  and  Development in  Mathematics and  Applications  (CIDMA)} \\ {\small  Campus  de  Santiago,  3810-183  Aveiro,  Portugal}
  }


\date{October 2021}




\maketitle


\begin{abstract} 
We study the time evolution of spherical, excited ($i.e.$ nodeful)  boson star models. We consider a model including quartic self-interactions, controlled by a coupling $\Lambda$. Performing non-linear simulations of the Einstein-(complex)-Klein-Gordon system, using as initial data equilibrium boson stars solutions of that system, we assess the impact of $\Lambda$ in the stability properties of the boson stars. In the absence of self-interactions ($\Lambda=0$), we observe the known behaviour that the excited stars in the (candidate) stable branch decay to a non-excited star without a node; however, we show that for large enough values of the self-interactions coupling, these excited stars do not decay (up to timescales of about $t\sim 10^4$). The stabilization of the excited states for large enough self-interactions is further supported by evidence that the nodeful states dynamically form through the gravitational cooling mechanism, starting from  dilute initial data. Our results support the healing power (against dynamical instabilities) of self-interactions, recently unveiled in the context of the non-axisymmetric instabilities of spinning boson stars.
\end{abstract}


\section{Introduction}
Elementary quantum mechanics describes how the hydrogen atom has different states, or orbitals, for an electron - see $e.g.$~\cite{Thaller}. There is, however, only one ground state, which is dynamically stable: the $1s$ orbital.  All other orbitals are excited states and if the electron is excited into one of the latter, it ends up decaying to the ground state, unless an external forcing is present. Some excited states ($ns$ orbitals, with $n>1$) are also spherical, just as the ground state, but, unlike the latter, the wave function has radial nodes.

Boson stars (BSs)~\cite{Kaup:1968zz,Ruffini:1969qy} are static, horizonless, localized solutions of the Einstein-(complex, massive) Klein-Gordon system - see also $e.g.$~\cite{Schunck:2003kk,Liebling:2012fv,Herdeiro:2017fhv,Herdeiro:2019mbz,Herdeiro:2020jzx}. There is some parallelism between these boson stars and atomic orbitals~\cite{Herdeiro:2020kvf}, but with some key differences. Firstly, whereas the Schr\"odinger orbitals emerge due to the  potential well created by the Coulomb potential of a nucleus with the opposite electric charge of the electron, boson stars create their own potential well, due to their self-gravity. This is possible due to the non-linearities of General Relativity. Secondly, whereas the Schr\"odinger orbitals are rightfully interpreted as a probability amplitude of a single (fermionic) particle, boson stars are superpositions of many bosons in the same state ($i.e.$ with the same frequency and multipolar distribution), so that a classical description is justified. Thirdly, each hydrogen orbital has a single frequency, whereas boson stars with a certain symmetry (say, spherical) have an interval range of possible frequencies. 

Similarly to atomic orbitals, there are different types of boson stars, one of which is the fundamental ground state, hence analogous to the $1s$ hydrogen orbital. This is the spherical solution for which the scalar field has no nodes; for a certain frequency range, hereafter dubbed the \textit{stable branch}, such boson stars are dynamically stable against small perturbations and remain dynamically robust in fully non-linear numerical evolutions~\cite{Seidel:1990jh}. Then, there are different types of excited states~\cite{Herdeiro:2020kvf}, the simplest of which are also spherical boson stars but for which the scalar field has radial nodes, which are counted by an excitation  integer number $n$, hence analogous to the  hydrogen $ns$-orbital ($n>1$). Such stars decay, when slightly perturbed, to the fundamental state (or black holes)~\cite{Balakrishna:1997ej}. The goal of this paper is to inquire whether  self-interactions can stabilize such excited ($i.e.$ nodeful) spherical boson stars. 

The motivation for the last question comes from recent studies of \textit{spinning} boson stars, where it was shown that a certain non-axisymmetric instability of these stars found in~\cite{sanchis2019nonlinear} can be mitigated~\cite{DiGiovanni:2020ror}, or even quenched~\cite{Siemonsen:2020hcg}, by introducing appropriate self-interactions in the scalar field model. Since spinning stars can be faced as an excited state, within the family of boson stars~\cite{Sanchis-Gual:2021edp}, where the excitation quantum number is $m\in \mathbb{Z}$, the azimuthal harmonic index, then, one may ask if,  by the same token, such self-interactions could improve the dynamical robustuness of nodeful excited states, for which the excitation number is $n\in \mathbb{N}$, the number of radial nodes.   

In this paper we provide numerical evidence that this is indeed the case, and that self-interactions can stabilize radially excited boson stars.  By performing numerical evolution of the fully non-linear  Einstein-(complex)-Klein-Gordon system, and using as initial data equilibrium boson stars solutions (with nodes) of that system, we assess the impact of  a parameter $\Lambda$, that controls the self-interactions, in the stability properties of the nodeful boson stars. In the absence of self-interactions ($\Lambda=0$), we recover the result first discussed in~\cite{Balakrishna:1997ej}  that the excited stars in the (candidate) stable branch decay to a non-excited star without a node. Cranking up $\Lambda$,  however, we observe that for large enough values these excited stars do not decay (up to timescales of about $t\sim 10^4$). 

To provide a different test to this stabilization mechanism of the excited stars, for large enough self-interactions, we also analyse the possible \textit{formation} of nodeful states dynamically, through the gravitational cooling mechanism, starting from  dilute initial data. Again we observe a sharp contrast depending on $\Lambda$. For $\Lambda=0$, nodeful initial data collapses by gravitational cooling to form a boson star that has no nodes; but for large $\Lambda$ we observe that the compact star that forms, albeit still undergoing an asymptotic relaxation process, retains the nodeful structure.  

This paper is organized as follows. In Section~\ref{sec2} we describe the excited states of spherical boson stars, as solutions of the appropriate Einstein-Klein-Gordon model.  In Section~\ref{section3} we discuss the framework for the numerical evolutions, and their main results both in the stability and formation scenarios. Finally, in Section~\ref{sec4} we provide our conclusions and final remarks.

\section{The model}
\label{sec2}

\subsection{The action and field equations}
We consider a model  with the following action, 
 for Einstein's gravity minimally coupled to a massive, complex, self-interacting scalar field  $\Phi $  
\begin{equation}
\label{action}
 \mathcal{S}=\int  d^4x\sqrt{-g}\left[
\frac{R }{16\pi}  
  -\frac{1}{2}  \partial_\mu\Phi^* \partial^\mu\Phi  -\frac{1}{2}   U(|\Phi|^2)
 \right] \ .
\end{equation} 

The resulting field equations are:
\begin{equation}
\label{E-eq}
 R_{\mu\nu}-\frac{1}{2}g_{\mu\nu}R=8 \pi ~T_{\mu\nu} \  , \qquad 
\Box \Phi =\frac{d U}{d |\Phi|^2}\Phi  \ ,  
\end{equation}
where 
\begin{equation}
T_{\mu\nu} =  
  \partial_{(\mu}\Phi^{ *}\partial_{\nu)}\Phi 
-\frac{1}{2}g_{\mu\nu} 
[  
 \partial_\alpha\Phi^{*}\partial^\alpha\Phi
+U(|\Phi|^2)
] \ ,
\end{equation}
is the
 energy-momentum tensor  of the scalar field.
In this work we choose the scalar field potential to have the following form
\begin{eqnarray}
U(|\Phi|^2)=\mu^2 |\Phi|^2+\frac{\lambda}{2} |\Phi|^4 \ ,
\end{eqnarray}
where $\mu$ is the scalar field mass and $\lambda$ is the parameter controling the quartic self-interactions. This model has been considered, in the context of boson stars, in a number of previous works, including~\cite{Colpi:1986ye,Herdeiro:2015tia,Escorihuela-Tomas:2017uac}.

\subsection{ The ansatz}
We are interested in spherical boson star solutions with radial nodes. These are described by static, spherically symmetric geometries and represent excited states of the fundamental (nodeless) spherical boson stars. The quartic term, moreover, generalizes the usual mini-boson stars, found for $\lambda=0$. 

The spacetime geometry is described by the following
metric ansatz in isotropic coordinates
\begin{eqnarray}
\label{metric}
ds^2=  -e^{2F_0(r )} dt^2+e^{2F_1(r )} 
\left(
       dr^2+r^2 \left[d\theta^2+\sin^2\theta  d\varphi^2\right]
\right) \        ,
\end{eqnarray}
in terms of two metric radial functions 	$F_{0,1}$.
Also,
$(r,\theta,\varphi)$
are spherical coordinates with the usual range (in particular $0\leqslant r<\infty$) and $t$ is the time coordinate.

The scalar field  $\Phi $
has the standard form for boson stars
\begin{eqnarray}
\label{ansatz-general}
\Phi =\phi(r )e^{-i \omega t }  \ ,
\end{eqnarray}
where $\omega>0$ is the field's oscillating frequency.
Here,   $\phi$ is the field's amplitude  and it 
is a real function. 
One can easily shown that the field ansatz (\ref{ansatz-general}) leads to a spherically symmetric energy-momentum tensor, 
thus compatible with the line element (\ref{metric}).

Below we shall sometimes use the areal radius ${\rm R}$, defined as ${\rm R}\equiv re^{F_1(r)}$, that has a clear geometrical  significance, and therefore it is a more meaningful coordinate when comparing radial profiles of different solutions.

\subsection{The  boundary conditions}

In this setting, obtaining boson star solutions
reduces to solving a set of three\footnote{There is also one constraint equation 
which is used to monitor the accuracy of the numerical results.}  
ordinary differential equations (ODEs),
for the metric functions $F_{0,1} $
and the scalar amplitude   $\phi $. 
These ODEs are subject to
the following set of boundary conditions.
At $r=0$ one imposes (from an expansion of the equations of motion near the origin),
\begin{eqnarray}
\partial_r F_{0,1}=0 \ , \qquad \partial_r \phi =0 \ .
 \end{eqnarray}
At infinity, asymptotic flatness requires that all functions vanish,
\begin{eqnarray}
 F_{0,1}= \phi =0 \ .
 \end{eqnarray} 

The numerical evaluation of the equations of motion (via a shooting method) is done in units with $\mu=1$,
such that the only input parameter of the model is $\lambda$.
Also, following \cite{Escorihuela-Tomas:2017uac},
we define 
\begin{eqnarray}
\Lambda:=\frac{\lambda}{4\pi} \ .
 \end{eqnarray}
In the remaining of this paper, $\Lambda$ is the parameter that shall be used to discuss the impact of the self-interactions.
 
\subsection{The  solutions }
Solving the equations of motion, a discrete set of families of solutions are found, label by $n$, the number of nodes of the scalar amplitude $\phi$. For $n=0$ we have the fundamental family - see $e.g$~\cite{Cunha:2017wao} for a description of its properties. Here, let us briefly discuss the properties of the first excited state - the one-node solutions, $n=1$, for different values of $\Lambda$.
 
\begin{figure}[h!]
\centering
\includegraphics[height=2.27in]{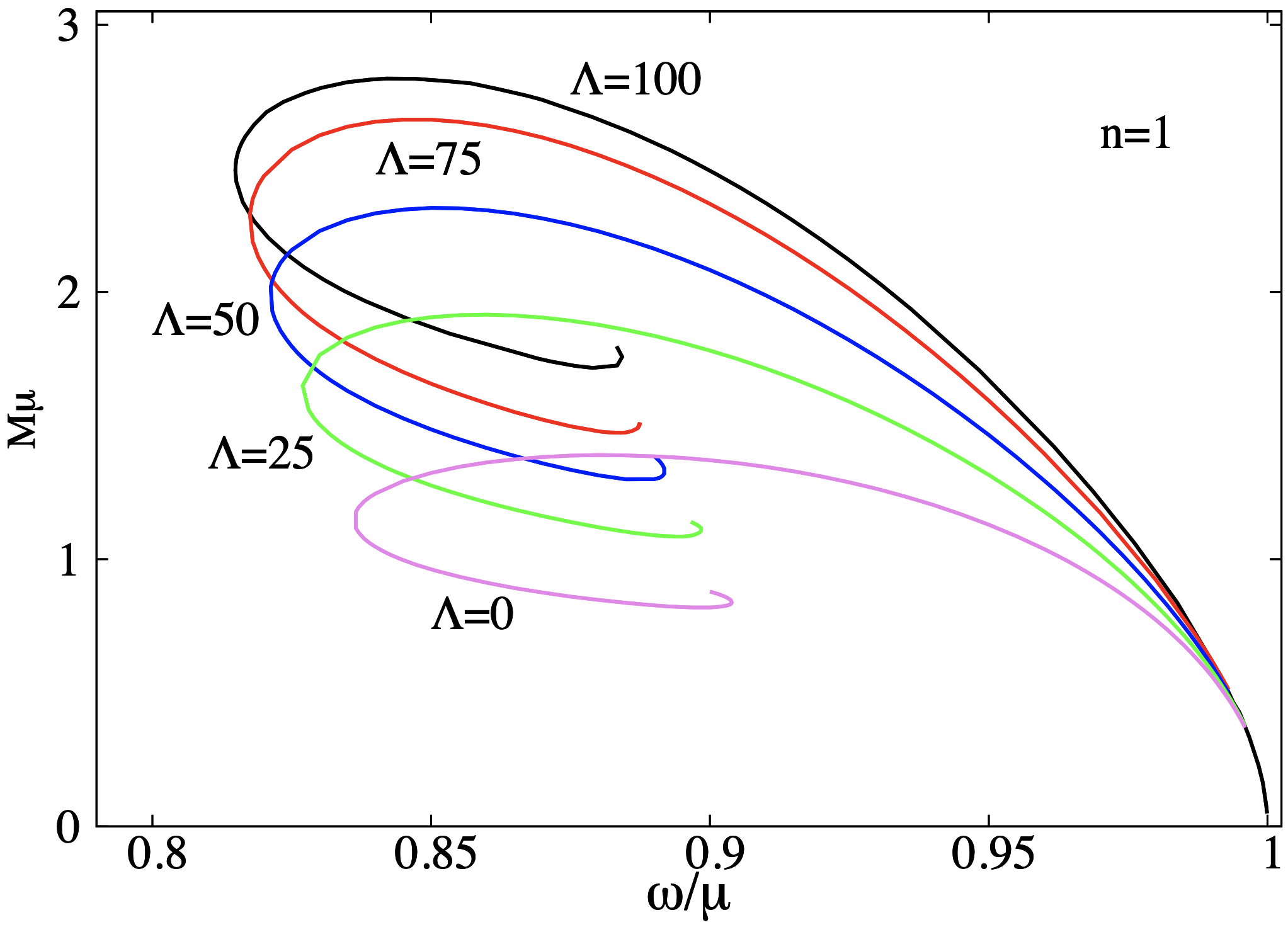}  
\includegraphics[height=2.3in]{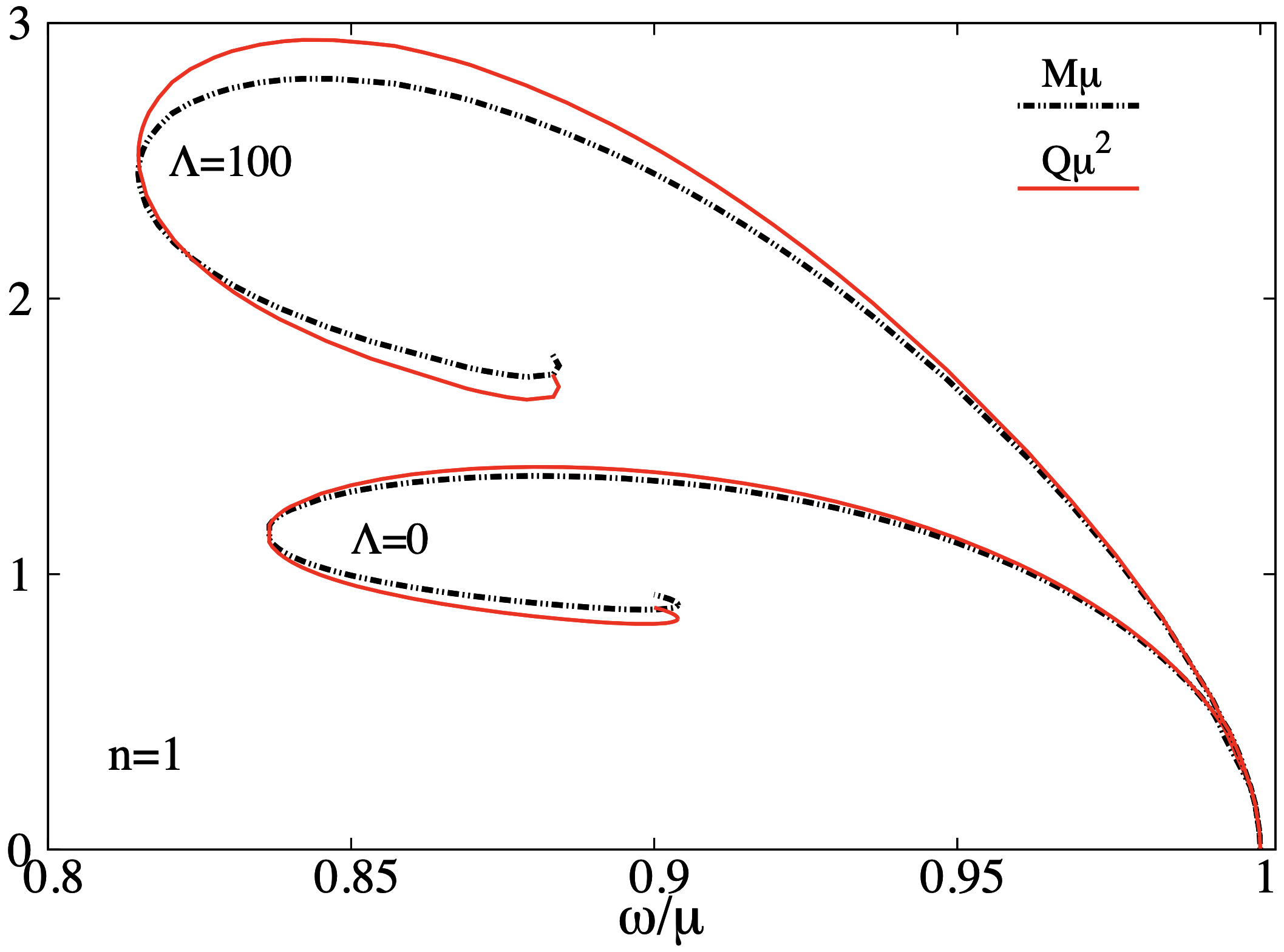}  
\caption{ The frequency $vs.$ mass diagram is shown for boson stars with  one node for the model with: (left panel)  $\Lambda=100,~75,~50,~25,~0$; (right panel)  $\Lambda=100,~0$,  showing also the  frequency $vs.$ Noether charge diagram.
}
\label{fig1}
\end{figure}

Such boson stars exist for $\omega_{\rm min}<\omega<\mu$, where the value $\omega_{\rm min} $ depends on the node number $n$ (with $\omega_{\rm min}\simeq 0.8149$ for $n=1$). Then, fixing $n$ and $\Lambda$, one finds that the solutions sit on a spiraling curve, when plotted in an ADM  
 mass ($M$) $vs.$ frequency ($\omega$) diagram. Some examples of this spiraling curve are displayed in Figure~\ref{fig1} (left panel), for different values of $\Lambda$.  
One observes that the picture familiar from the fundamental 
spherically symmetric mini-boson stars  is recovered for general $\Lambda$, with 
 the $(\omega,M)$ 
curve spiraling towards a central region of the diagram where the numerics become increasingly challenging. In the case of the fundamental $n=0$ boson stars, the \textit{stable branch} is the one between the maximal frequency $\omega=\mu$ and the maximum of the ADM mass. By analogy we shall call the \textit{candidate stable branch} the corresponding region for these solutions with $n=1$. In the right panel of  Figure~\ref{fig1} we can see both the ADM mass $vs.$ frequency curve as well as the Noether charge $vs.$ frequency curve for the illustrative cases of
 $\Lambda=0,100$. The Noether charge $Q$ is a global charge due to the $U(1)$ global symmetry of the complex scalar field model, and it can be interpreted as a measure of the number of scalar particles in the boson star (a notion made precise upon quantization). If $Q<M\mu$ this means that the star has excess energy (rather than binding energy), and therefore the stars become energetically unstable against fission - see $e.g.$~\cite{Cunha:2017wao}. The plot shows this only occurs, when moving along the spiral starting from the maximal frequency, near the minimum frequency, and therefore away from the candidate stable branch.

\begin{figure}[h!]
\centering
\includegraphics[height=2.27in]{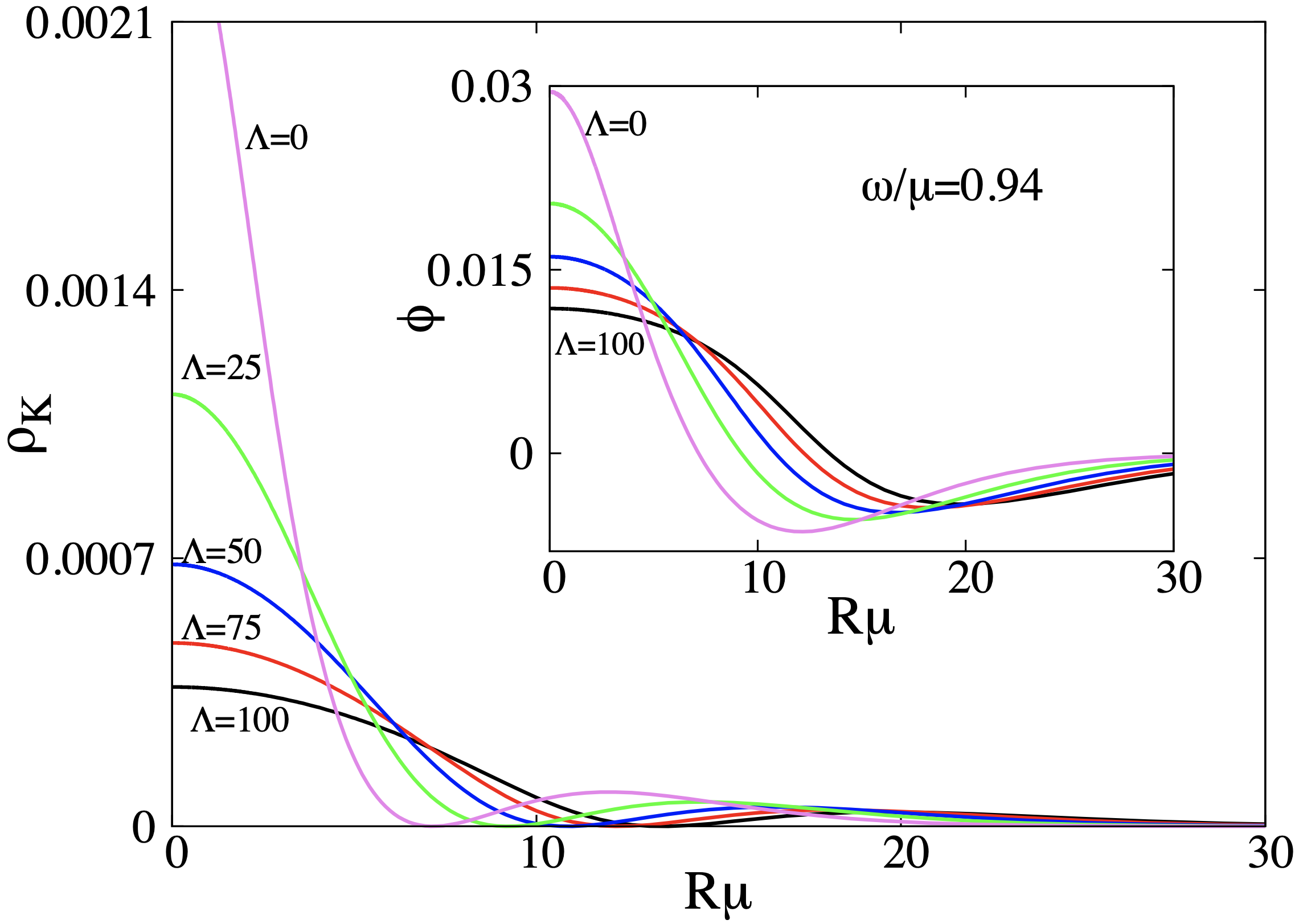}  
\includegraphics[height=2.3in]{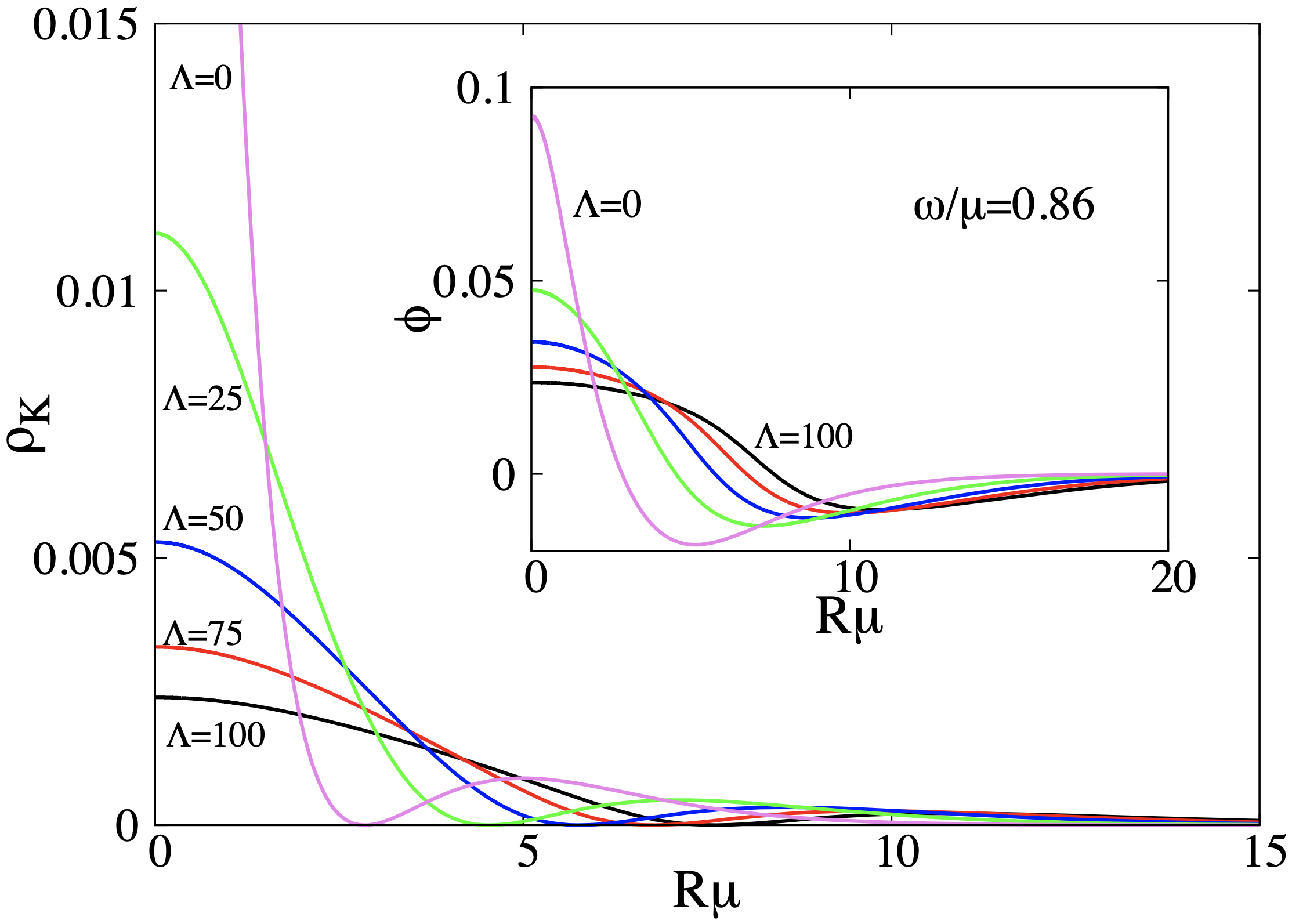}  
\caption{ Radial profile (in terms of the areal radius) of the Komar energy density/scalar field amplitude (main panels/insets)  for $\Lambda=0, 25, 50, 75, 100$  and $\omega=0.94$ (top left panel) or $\omega=0.86$ (top right panel).
}
\label{fig12}
\end{figure}

In the dynamical analyses of the next section we shall be interested in examining sequences of solutions for fixed $\omega$ and varying $\Lambda$. In Figure~\ref{fig12} we show the radial profile of the scalar field amplitude for $\Lambda=0, 25, 50, 75, 100$ and $\omega=0.94$ (inset left panel) and $\omega=0.86$ (inset right panel). One observes that: $i)$ the solutions always have one radial node; $ii)$ for fixed $\omega$, increasing $\Lambda$, the radial node occurs at a larger $r$ coordinate; $iii)$ for fixed $\omega$, increasing $\Lambda$, the amplitude of values of $\phi$ between its maximum and minimum, $\Delta \phi=\phi_{\rm max}-\phi_{\rm min}$  decreases; $iv)$ for fixed $\Lambda$, decreasing $\omega$, the solutions become more "compact", with the node occurring at smaller $r$ and with a higher $\Delta \phi$.

One can also analyse the Komar energy density profiles, for the same models as in the previous paragraph. Since for a stationary, asymptotically flat spacetime admitting an everywhere timelike Killing vector field $k=\partial_t$ the ADM mass $M$ coincides with the Komar mass at infinity, we can write the standard Komar mass expression as
\begin{equation}
M=-\frac{1}{8\pi}\oint dS_{\alpha\beta}D^\alpha k^\beta=\int drd\theta d\varphi (T-2T_t^t)\sqrt{-g} \ ,
\end{equation}
where $T\equiv T^\alpha_\alpha$. The radial profile of Komar energy density, 
\begin{equation}
\rho_K := T-2T_t^t \ ,
\label{ked}
\end{equation}
is exhibited in Figure~\ref{fig12} (main panels) leading to similar conclusions as the analysis of the scalar field amplitude. 

\begin{figure}[h!]
\centering
\includegraphics[height=2.27in]{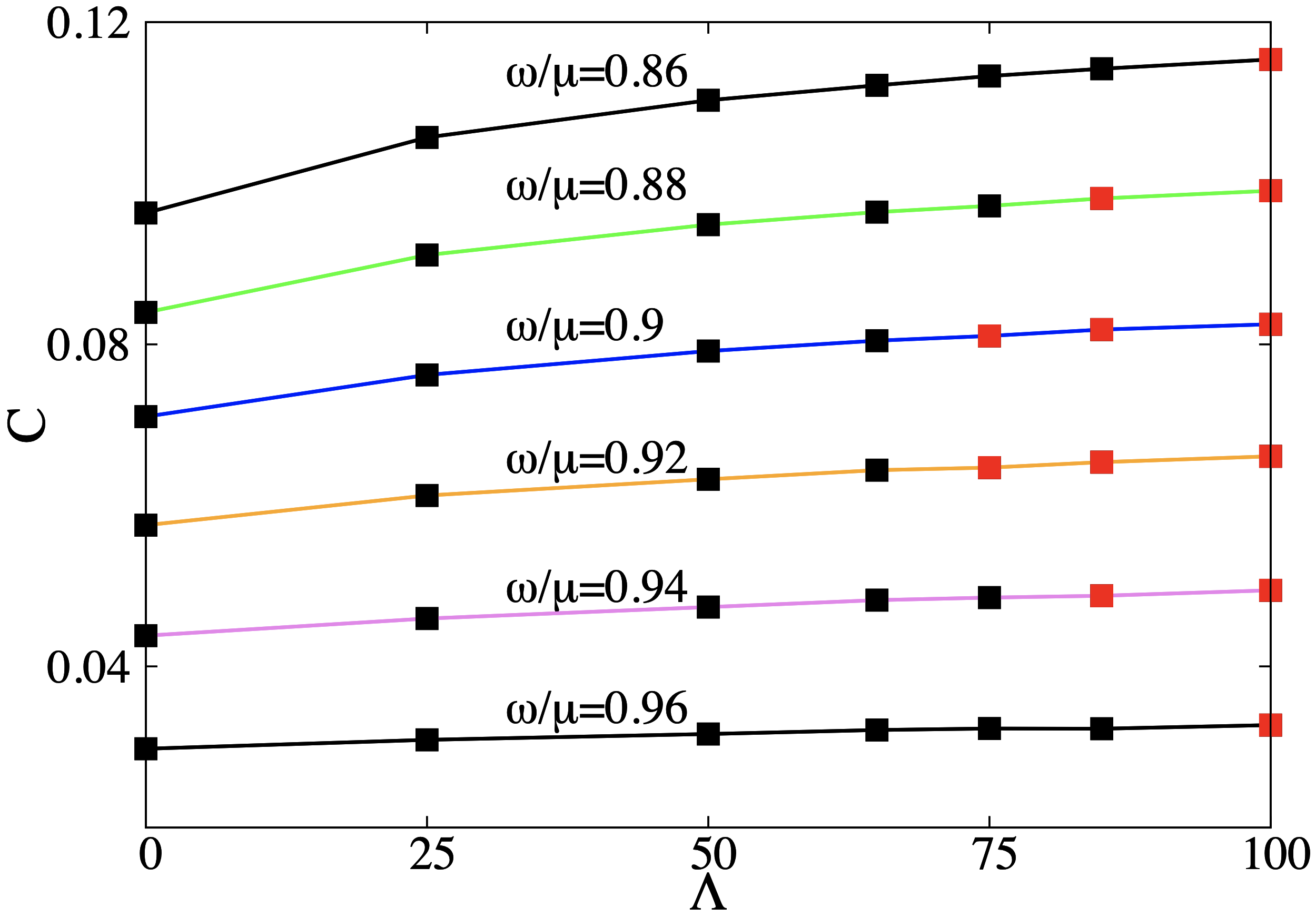}  
\caption{ Compactness as a function of $\Lambda$ and $\omega$. The black (red) points correspond to the solutions for which an instability (no instability) is seen in the evolutions of the next Section.
}
\label{fig13}
\end{figure}

Finally, in Figure~\ref{fig13} we exhibit the compactness of the boson stars, defined as
\begin{equation}
C := \frac{2M_{99}}{{\rm R}_{99}} \ ,
\end{equation}
where ${\rm R}_{99}$ is the areal radius ${\rm R}$ that encloses $99\%$ of mass of the boson star. One can appreciate that: $i)$ for fixed $\Lambda$ the compactness increases when decreasing $\omega$ (within the candidate stable branch); $ii)$  for fixed $\omega$ the compactness increases when increasing $\Lambda$. Thus, increasing the self-interactions makes the boson stars more compact in the sequences of fixed $\omega$ solutions analysed in the next Section.

\section{Numerical evolutions}
\label{section3}

Having constructed the $n=1$ spherical boson stars of the model defined by the action~\eqref{action} we shall proceed to analyse their dynamical stability. Thus, 
in this section we discuss the non-linear evolutions of the static models described above and the possible dynamical formation of these excited boson stars.

\subsection{Basic equations}
\label{sec:formalism}

 The spacetime metric can be written as
\begin{equation}
ds^{2}=g_{\alpha\beta}dx^{\alpha}dx^{\beta}=-\alpha^{2}dt^2+\gamma_{ij}(dx^{i}+\beta^{i}dt)(dx^{j}+\beta^{j}dt),
\end{equation}
where $\alpha$ and $\beta^{i}$ are the lapse and shift functions, respectively, and $\gamma_{ij}$ the spatial metric. For numerical evolutions, the spatial line element in spherical symmetry is given by
\begin{equation}\label{isotropic}
 dl^2 = e^{4\chi } \left[ a(t,r)dr^2+ r^2\,b(t,r)  d\Omega^2 \right] \ ,
\end{equation}
where $d\Omega^2 = d\theta^2 + \sin^2\theta \,d\varphi^2$
and $a(t,r)$ and $b(t,r)$ are two non-vanishing conformal metric functions. They are related to physical metric by the conformal decomposition $\gamma_{ij}=e^{4\chi}\hat\gamma_{ij}$, with $e^\chi = (\gamma/\hat \gamma)^{1/12}$, where $\chi$ is the conformal exponent, and with $\gamma$
and $\hat\gamma$ being the determinants of the physical and conformal 3-metrics, respectively.  We use the Baumgarte-Shapiro-Shibata-Nakamura (BSSN) formalism to solve Einstein's equations~\cite{Baumgarte98,Shibata95} (see \cite{Alcubierre:2010is,Montero:2012yr,Sanchis-Gual:2015bh} for further details). The explicit form of the evolution equations for the gravitational field can be found in Eqs.~(9)-(11) and (13)-(15) in Ref.~\cite{Sanchis-Gual:2015bh}.

As in previous works~\cite{escorihuela2017quasistationary,Sanchis-Gual:2016tcm}, we solve the Klein-Gordon equation by introducing first-order variables defined as
\begin{eqnarray}
\Pi &:=& n^{\alpha}\partial_{\alpha}\phi=\frac{1}{\alpha}(\partial_{t}\phi-\beta^{r}\partial_{r}\phi) \ ,\\
\Psi&:=&\partial_{r}\phi \ .
\end{eqnarray}
with $n^{\alpha}=(1/\alpha,-\beta^{i}/\alpha)$, the future-pointing unit
normal vector to the spatial hypersurfaces of the spacetime foliation. We obtain the following system of first-order equations for the scalar field:
\begin{eqnarray}
\partial_{t}\phi&=&\beta^{r}\partial_{r}\phi+\alpha\Pi \ ,\\
\partial_{t}\Pi&=&\beta^{r}\partial_{r}\Pi+\frac{\alpha}{ae^{4\chi}}[\partial_{r}\Psi
+\Psi\biggl(\frac{2}{r}-\frac{\partial_{r}a}{2a}+\frac{\partial_{r}b}{b}+2\partial_{r}\chi\biggl)\biggl]
+\frac{\Psi}{ae^{4\chi}}\partial_{r}\alpha+\alpha K\Pi - \alpha \bigl(\mu^{2}+\lambda \phi^2\bigl)\phi \ ,
\label{eq:sist-KG}
\end{eqnarray}
where $K$ is the trace of the extrinsic
curvature, $K_{ij}$. The matter source terms contained in the right-hand sides of the gravitational field evolution equations, denoted by $\mathcal{E}$, $S_{a}$, $S_{b}$ and $j_r$, are components of the energy-momentum tensor and are given by
\begin{eqnarray}
\mathcal{E}&:=&n^{\alpha}n^{\beta}T_{\alpha\beta}=\frac{1}{2}\biggl(|\Pi|^{2}+\frac{|\Psi|^{2}}{ae^{4\chi}}\biggl) +\frac{1}{2}\mu^{2}|\phi|^{2}+\frac{1}{4}\lambda\,|\phi|^{4} \label{eq:rho}, \label{scalared}\\
j_{r}&:=&-\gamma^{\alpha}_{r}n^{\beta}T_{\alpha\beta}=-\frac{1}{2}\biggl(\Pi^{*}\Psi+\Psi^{*}\Pi\biggl)\ ,\\
S_{a}&:=&T^{r}_{r}=\frac{1}{2}\biggl(|\Pi|^{2}+\frac{|\Psi|^{2}}{ae^{4\chi}}\biggl) 
-\frac{1}{2}\mu^{2}|\phi|^{2}-\frac{1}{4}\lambda\,|\phi|^{4} \ ,\\
S_{b}&:=&T^{\theta}_{\theta}=\frac{1}{2}\biggl(|\Pi|^{2}-\frac{|\Psi|^{2}}{ae^{4\chi}}\biggl) -\frac{1}{2}\mu^{2}|\phi|^{2}-\frac{1}{4}\lambda\,|\phi|^{4}\ .
\end{eqnarray}

Finally, the Hamiltonian and momentum constraints are given by the following two equations: 
\begin{eqnarray}
\mathcal{H}&\equiv& R-(A^{2}_{a}+2A_{b}^{2})+\frac{2}{3}K^{2}-16\pi \mathcal{E}=0 \ ,\label{hamiltonian}\\
\mathcal{M}_{r}&\equiv&\partial_{r}A_{a}-\frac{2}{3}\partial_{r}K+6A_{a}\partial_{r}\chi
+(A_{a}-A_{b})\biggl(\frac{2}{r}+\frac{\partial_{r}b}{b}\biggl)-8\pi j_{r}=0 \ ,\label{momentum}
\end{eqnarray}
where $A_{ij}$ is the traceless part of the
conformal extrinsic curvature with $A_{a}\equiv A^{r}_{r}$ and $A_{b}\equiv A^{\theta}_{\theta}$.

\begin{figure}[t]
 \includegraphics[width=0.5\textwidth]{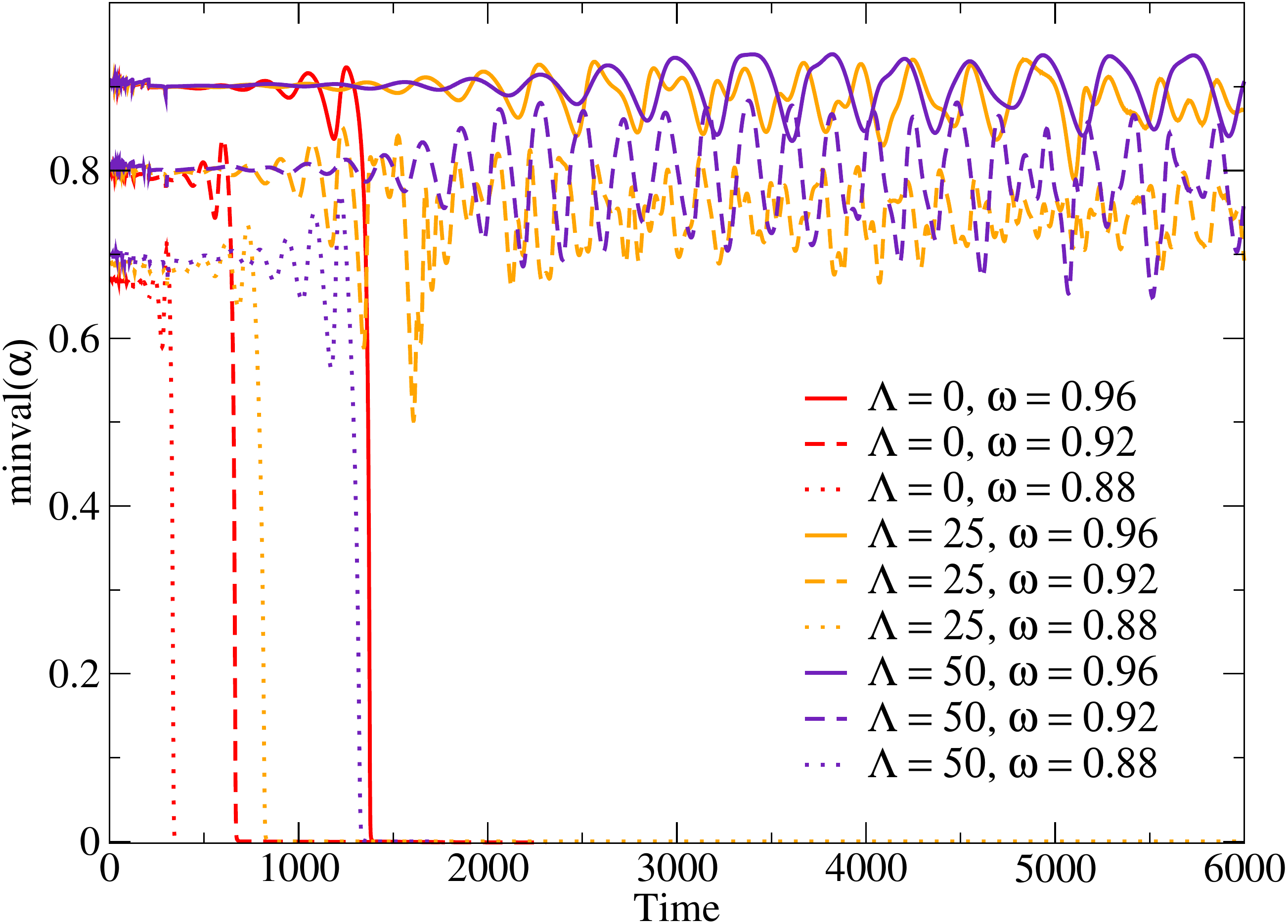} 
  \includegraphics[width=0.5\textwidth]{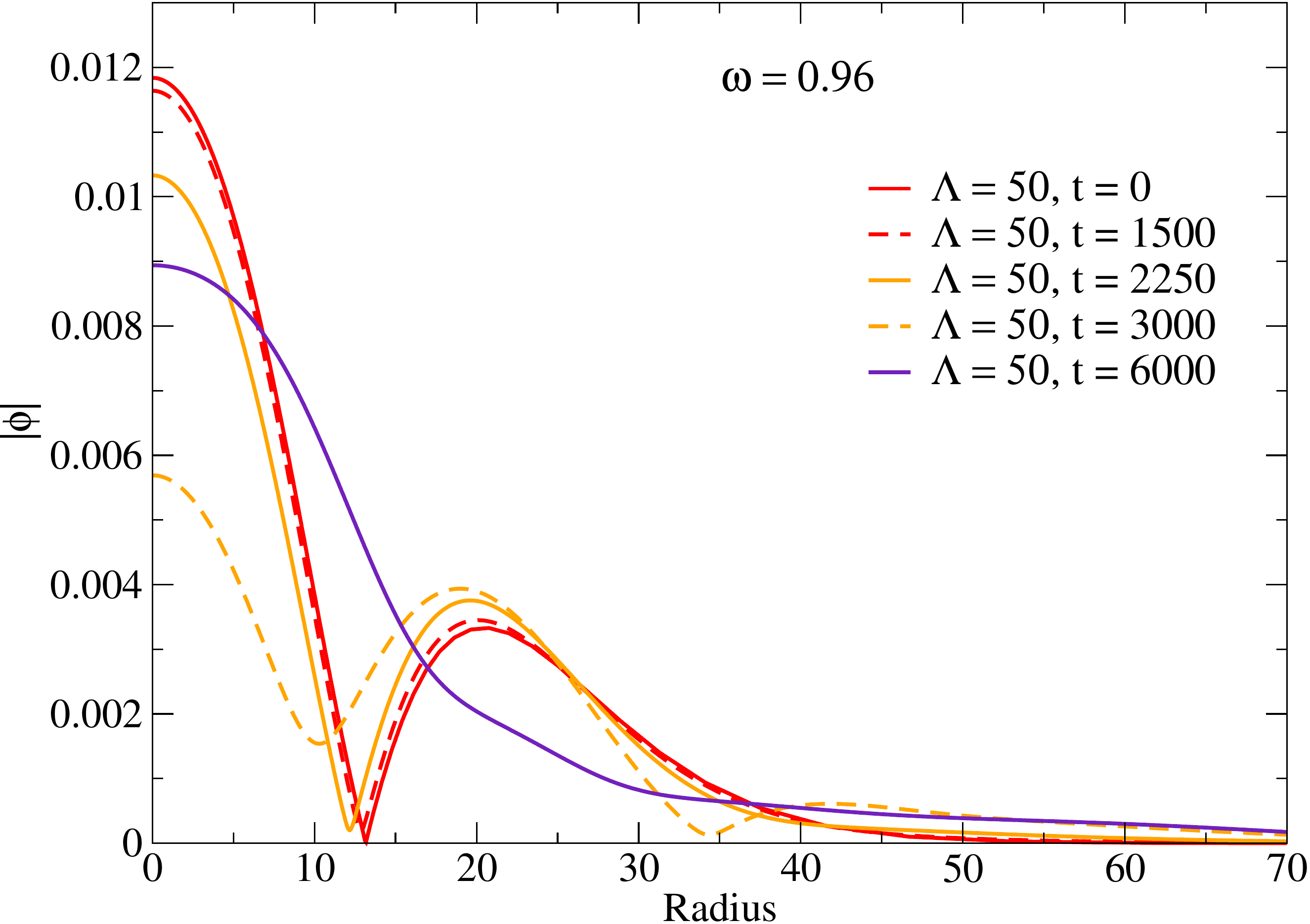}
   \includegraphics[width=0.5\textwidth]{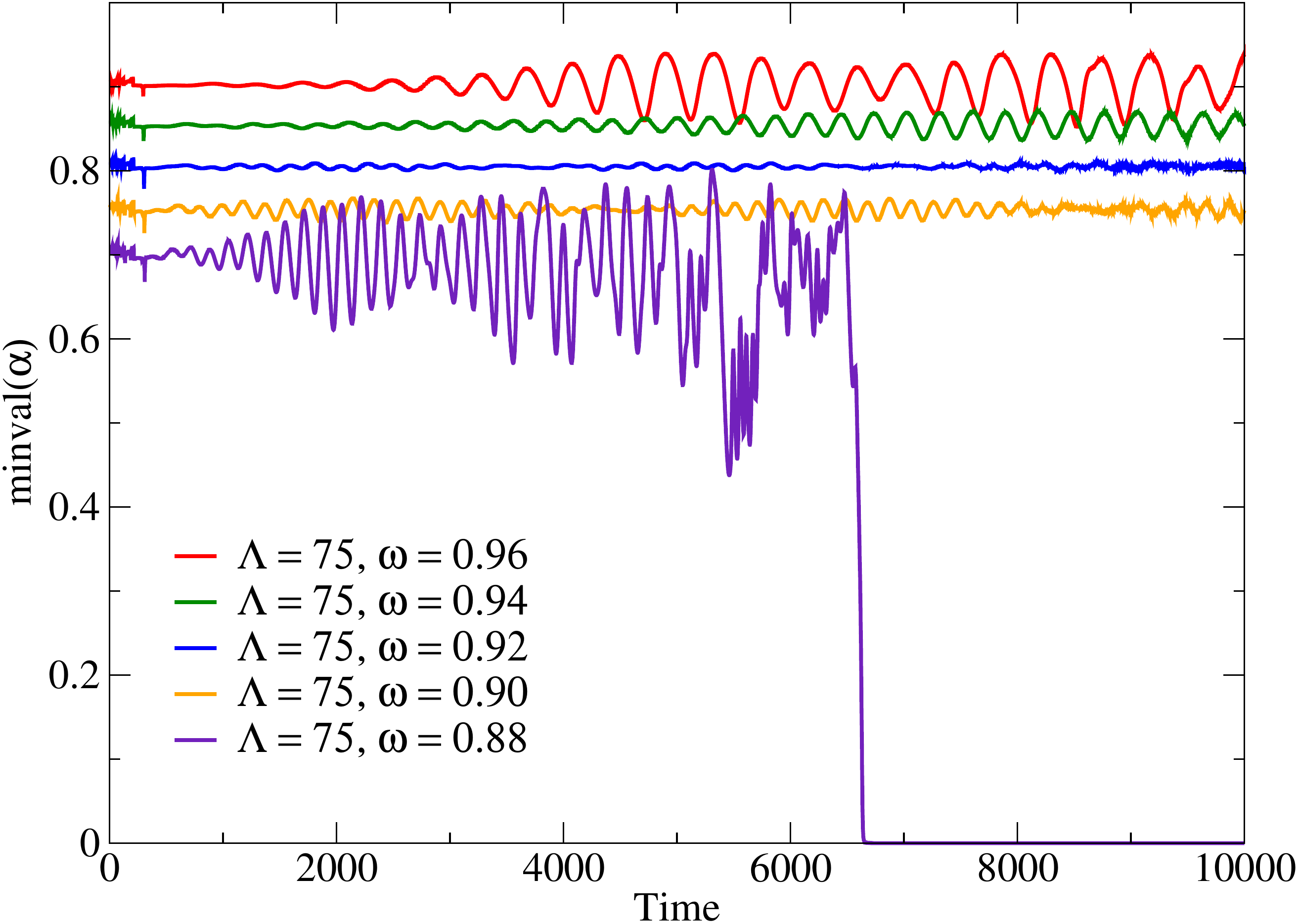} 
  \includegraphics[width=0.5\textwidth]{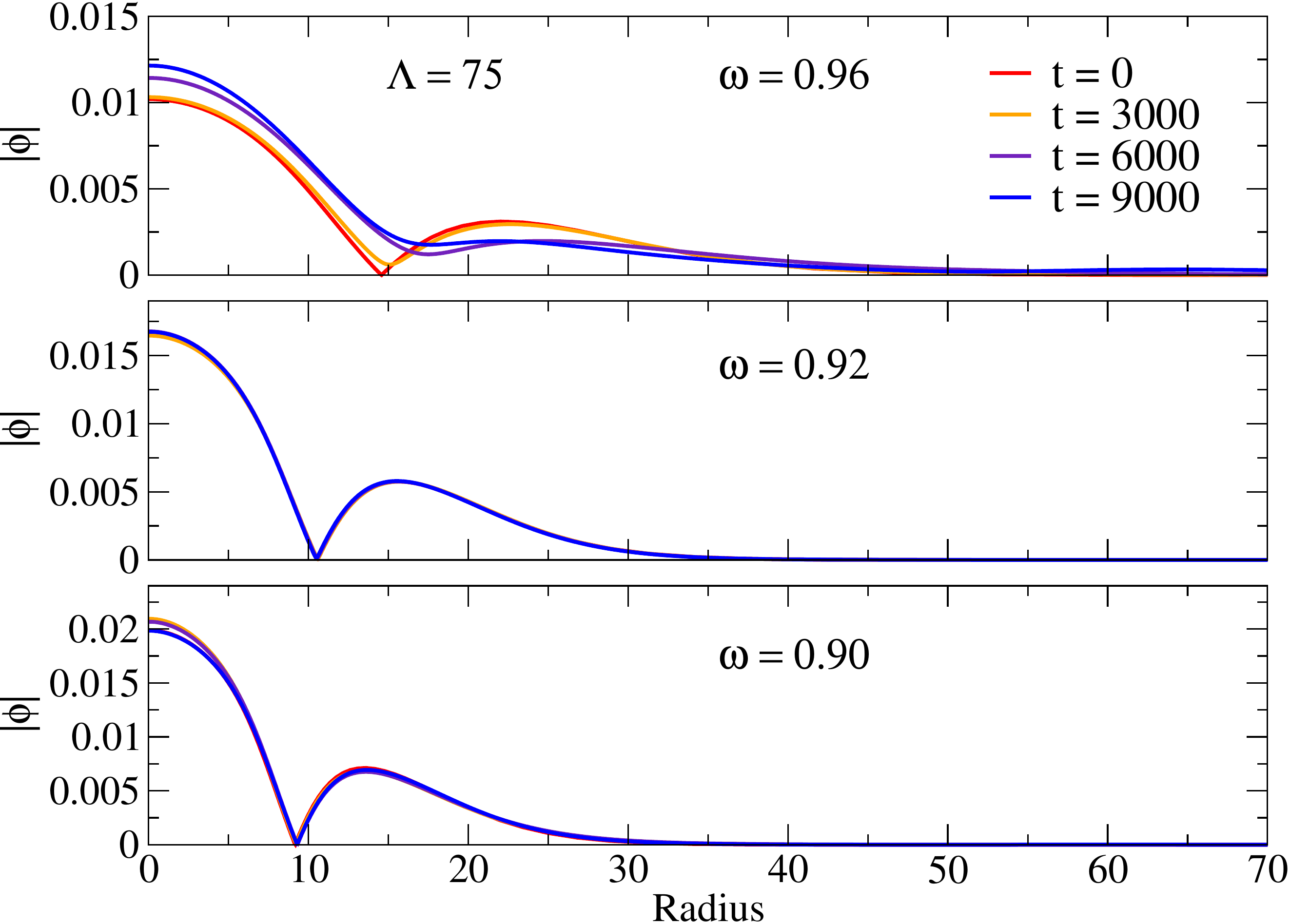}
\caption{(Top left panel) Time evolution of the minimum value of the lapse for different models with $\Lambda=\lbrace0, 25, 50\rbrace$ and $\omega=0.96, 0.92, 0.88$. (Top right panel) Radial profile of the scalar field magnitude at different times for the model with $\omega = 0.96$ and $\Lambda=75$. (Bottom left panel) Time evolution of the minimum value of the lapse for models with $\Lambda=75$.  (Bottom right panel) Radial profile of the scalar field magnitude at different times for three models with $\Lambda=75$.}
\label{fig:lambda}
\end{figure}

\subsection{Stability}
We solve the above equations using a numerical-relativity code that assumes spherical symmetry and spherical coordinates, described in~\cite{Montero:2012yr,Sanchis-Gual:2015bh,Sanchis-Gual:2015lje,escorihuela2017quasistationary,di2020dynamical,DiGiovanni:2021vlu}. The BSSN and Klein-Gordon coupled equations are solved using a second-order Partially Implicity Runge-Kutta scheme~\cite{Casas:2014,Isabel:2012arx}. The evolutions are performed in a logarithmic grid described in~\cite{Sanchis-Gual:2015sxa}, with a maximum resolution of $\Delta r=0.05$, a time step of $\Delta t = 0.3 \Delta r$, and the outer boundary placed at $r_{\rm{max}}=10000$.

We study the non-linear stability of excited boson star solutions with different values of the coupling, namely $\Lambda=\lbrace0,25,50,65,75,85,100\rbrace$. The spirals corresponding to $\Lambda=\lbrace 65,85\rbrace$ are not explicitly shown in Figure~\ref{fig1} (left panel), but one may easily guess their approximate location by interpolation.

The mini-boson star case ($\Lambda=0$) has been extensively investigated and found to be unstable (for the excited states) even in the candidate stable branch, which occurs before the maximum ADM mass model~\cite{Balakrishna:1997ej}. Numerical evolutions showed that the excited states decay to the fundamental solution without nodes. Since the maximum mass of the excited states is larger than that of the fundamental mini-boson stars, a large part of the sequence of equilibrium configurations collapse to a black hole after decaying. This behaviour is shown in the top left panel of Figure~\ref{fig:lambda}, where we plot the time evolution of the minimum value of the lapse, $\alpha$, for three different values of $\Lambda=\lbrace0,25,50\rbrace$ and a sample of three different frequencies ($\omega=0.88, 0.92, 0.96$). The smallest mass of the models with $\Lambda=0$ we have considered is $M=1.024$, corresponding to $\omega=0.96$, while the maximum mass of the mini-boson stars is $M_{\rm{max}}=0.633$; therefore, taking into account that in the collapse little mass is dispersed away, all these configurations must collapse to a black hole (cannot become a fundamental boson star). Indeed this is verified in  the top left panel of Figure~\ref{fig:lambda}, which is diagnosed from the fact the minimum value of the lapse  becomes zero for all models with $\Lambda=0$ (red lines).

Increasing the value of $\Lambda$ to 25 or 50 leads to another possible outcome for the same value of $\omega$. Now, only the most compact model (with the highest mass), corresponding to $\omega=0.88$ collapses to a black hole. But the two models with less mass and less compactness ($\omega=0.92, 0.96$), do not; rather the minimum value of the lapse oscillates around a non-zero value, diagnosing a relaxation to a new equilibrium state. This is possible because  the maximum mass of the fundamental branch also increases with $\Lambda$. The fate of the models that do not collapse into a black hole, is illustrated in the top right panel of Fig.~\ref{fig:lambda}, for the case with $\Lambda=50$ and $\omega=0.96$, where we show the radial profile of the scalar field magnitude, $|\phi|=\sqrt{\rm{Re}(\phi)^2+ \rm{Im}(\phi)^2}$, at different times. One observers that at the end of the simulation (indigo line) there is no node in the radial profile, confirming that the evolution led to the decay of the excited (nodeful) star to a fundamental (nodeless) star.

\begin{figure}[h!]
 \includegraphics[width=0.5\textwidth]{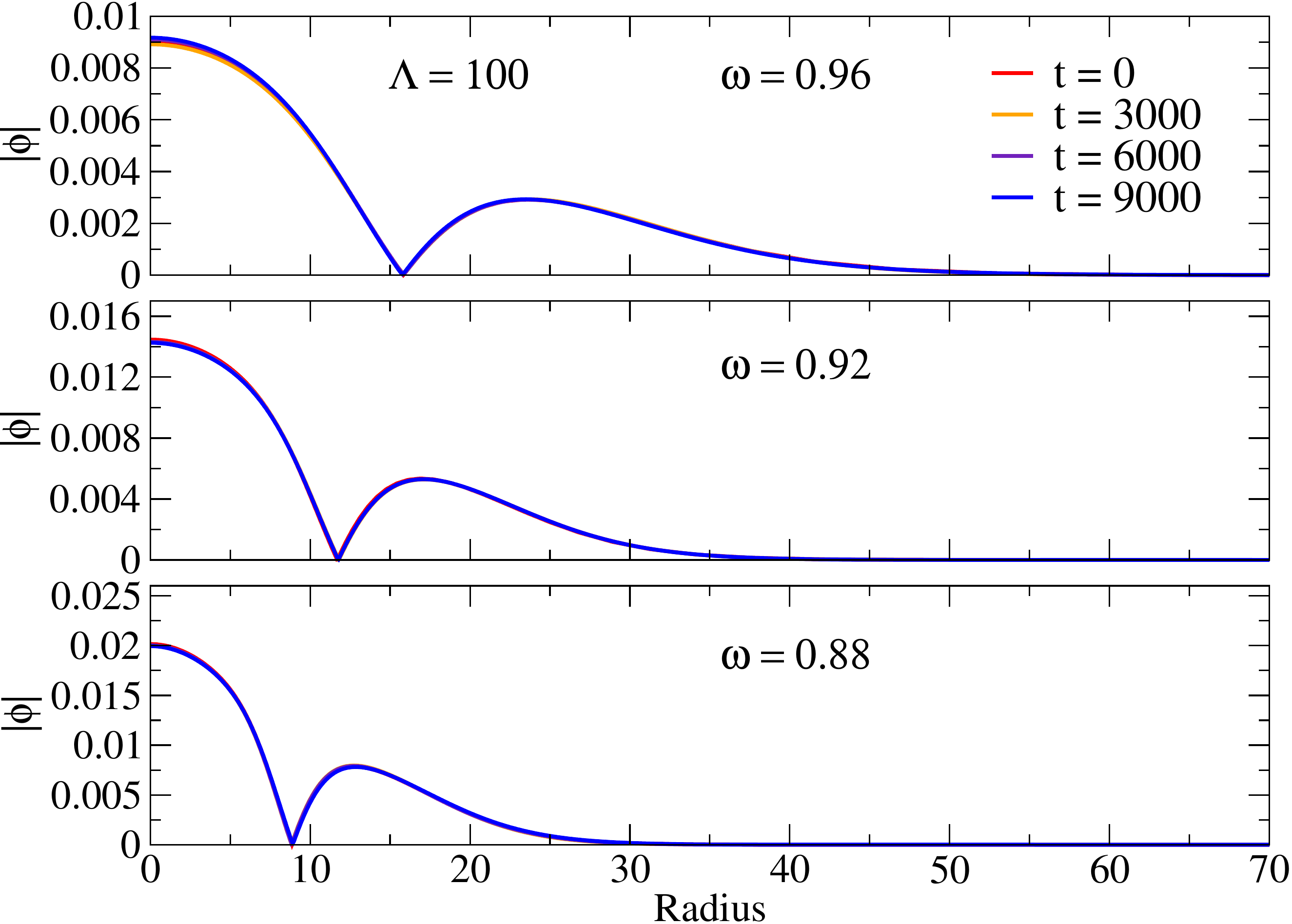} 
 \includegraphics[width=0.51\textwidth]{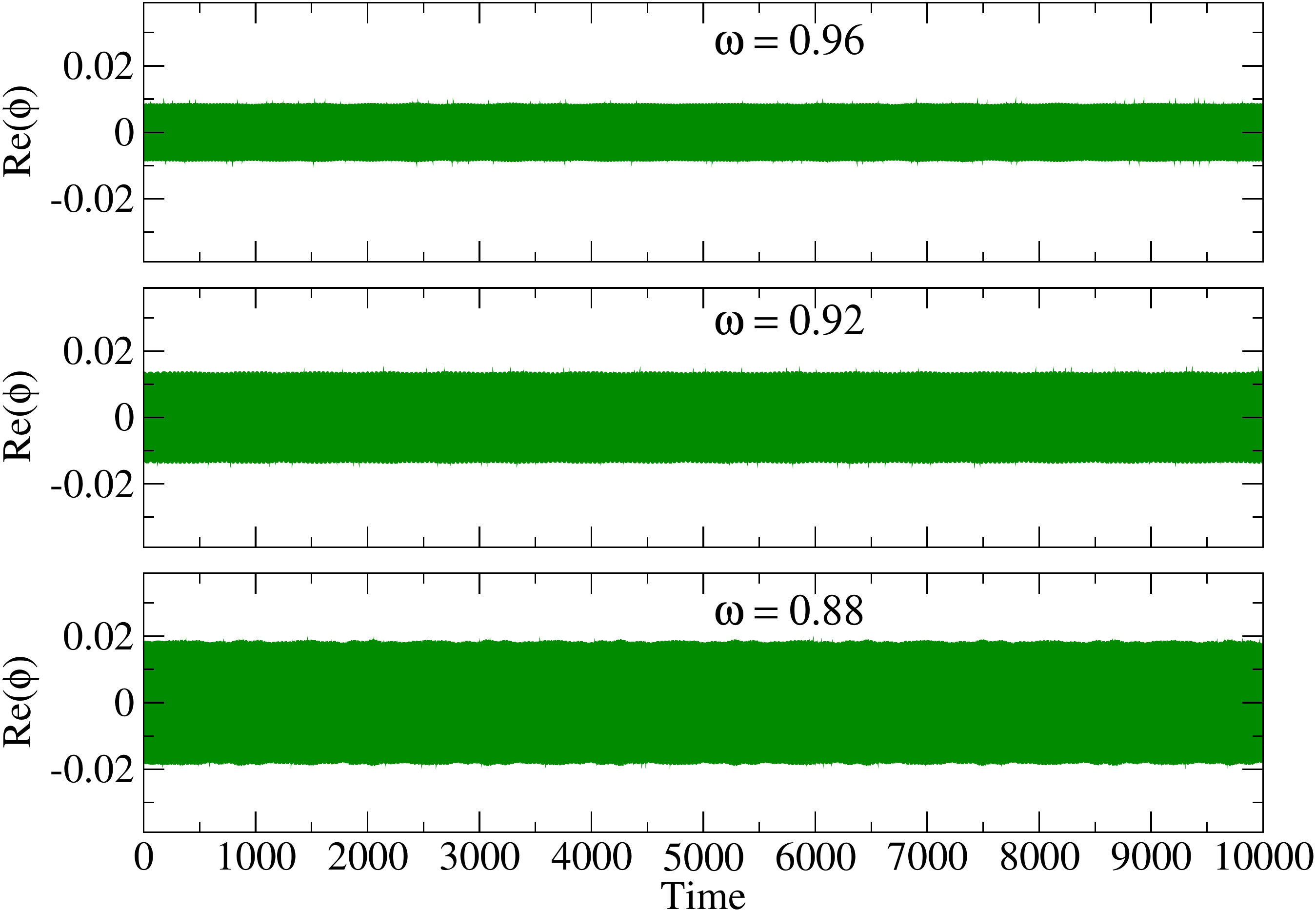}
  \includegraphics[width=0.5\textwidth]{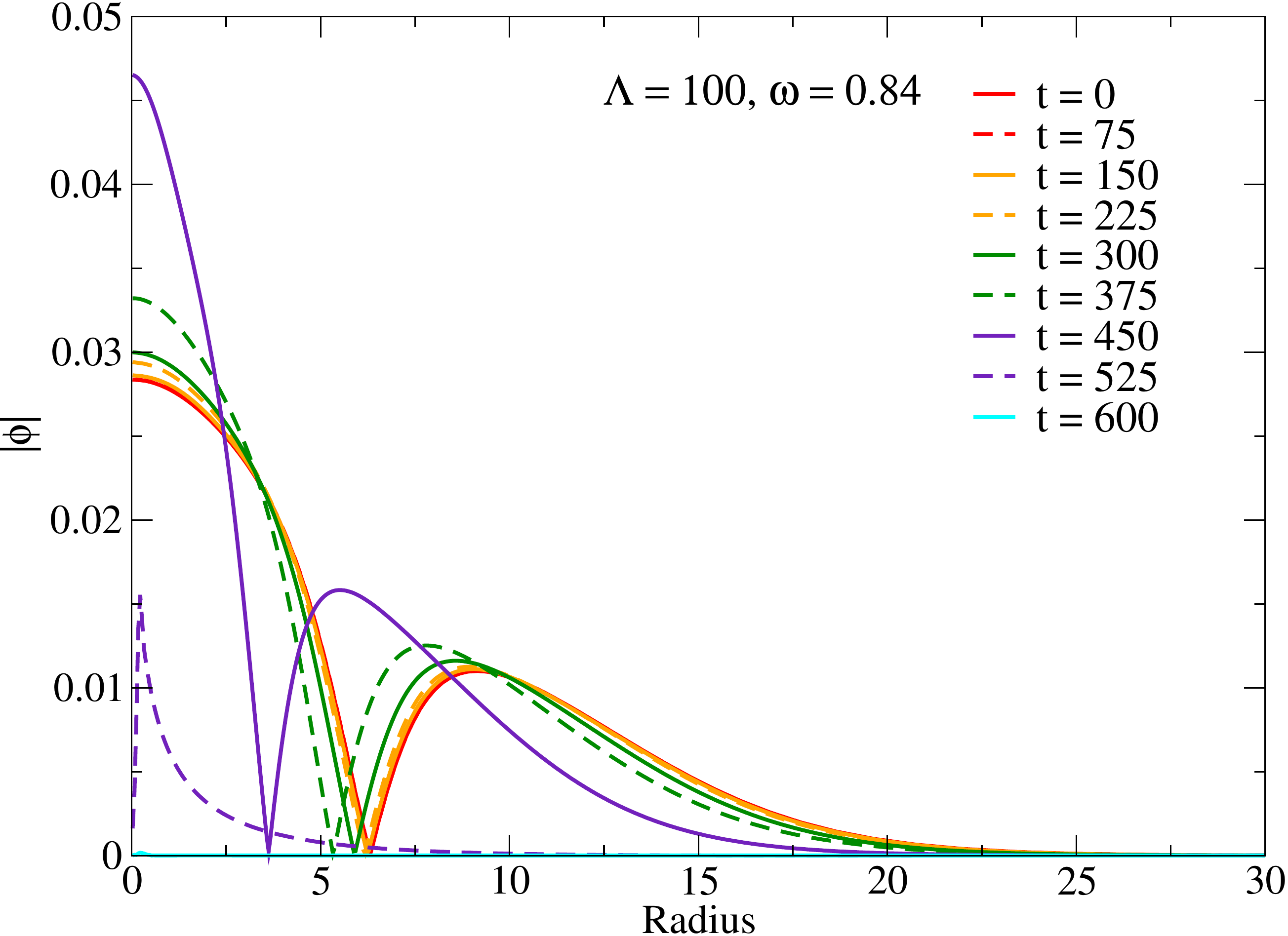}
  \includegraphics[width=0.5\textwidth]{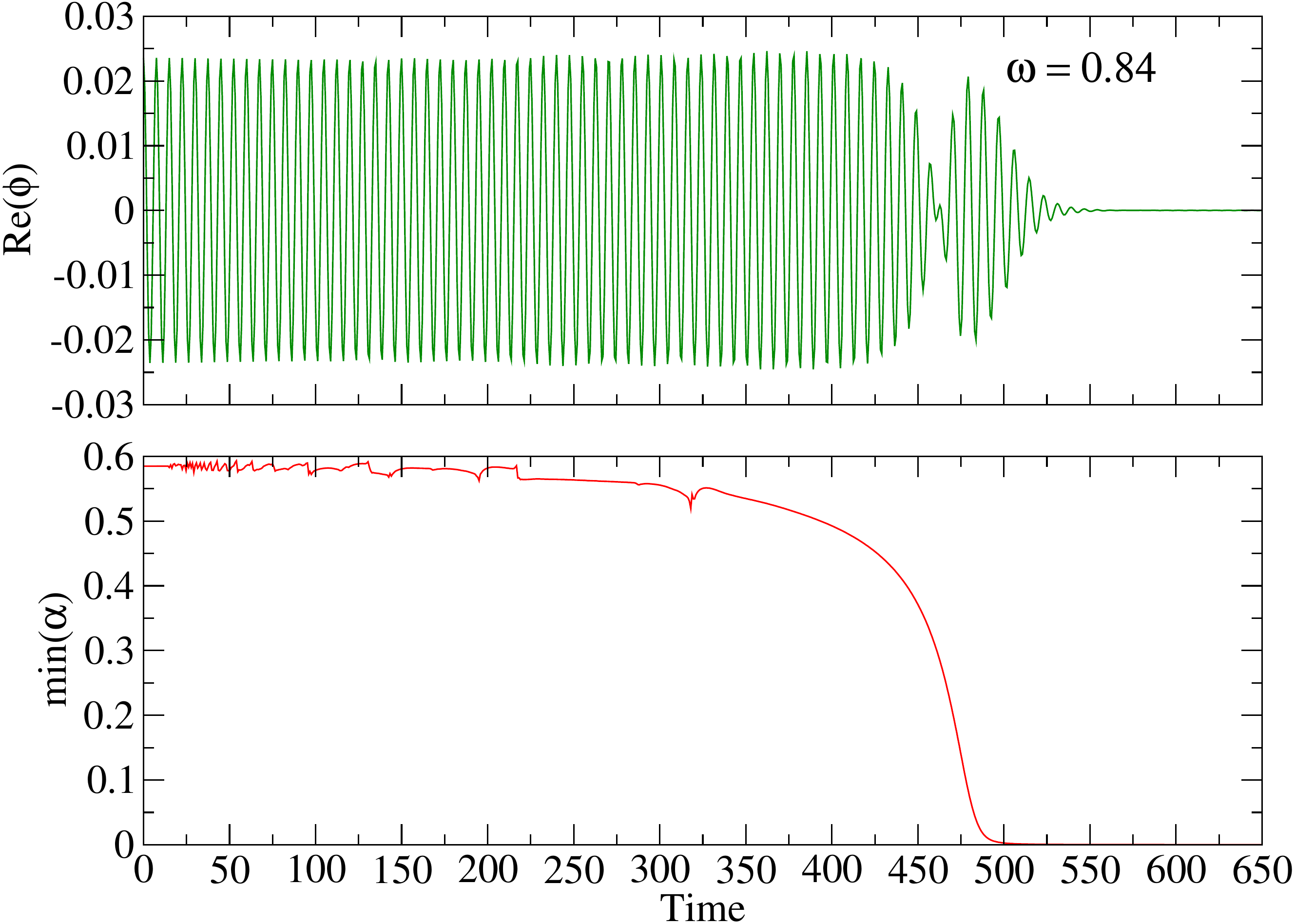}
\caption{(Top left panel) Radial profile of the scalar field magnitude $|\phi|$ at different times for three different models with $\Lambda=100$. (Top right panel) Time evolution of the amplitude of the real part of the scalar field extracted at $r=3$ for three different models. (Bottom left panel) Radial profile of the scalar field magnitude at different times for the unstable model with $\omega = 0.84$. (Bottom right panel) Time evolution of the amplitude of the real part of the scalar field extracted at $r=3$ and the minimum value of the lapse for the model with $\omega=0.84$.}
\label{fig:l100}
\end{figure}

A qualitatively different behaviour arises when we further increase the self-interactions parameter to $\Lambda=75$. Whereas low-compactness models, with $\omega\geqslant 0.96, 0.94$, are still unstable and quickly decay to the nodeless solution (see bottom left and right panels of Figure~\ref{fig:lambda}), fairly compact models with lower $\omega$ ($\omega\geqslant 0.92, 0.90$) become stable at least up to $t=10000$; even when explicitly perturbed, such excited solutions remain stable. There is, however,  an additional turning point around $\omega\geqslant0.88$, which is also unstable, losing the node and collapsing to a black hole. We emphasise all these solutions are in the candidate stable branch, before the maximum mass of the model.

Increasing further the self-interaction coupling to $\Lambda=100$ we find that all models in the candidate stable branch remain in the excited state. This is illustrated  in the top panels of Figure~\ref{fig:l100} , where the time evolution of the radial profile of $|\phi|$ and the amplitude of the real part of the scalar field for three values of $\omega$ is exhibited.

These results indicate that the addition of the self-interaction term can quench the decay to the fundamental branch, complementing previous findings on the stability of rotating boson stars with self-interaction potentials. Moreover, it is interesting to point out the correlation between the self-interaction potentials and the stabilization mechanism described in~\cite{Siemonsen:2020hcg}. 

For completeness, we have also evolved a couple of models in the unstable branch (beyond the maximum ADM mass) with $\Lambda=100$ (see bottom panels of Figure~\ref{fig:l100}). As expected these stars collapse to a black hole, but during the process \textit{they do not} lose the node. Therefore, the collapse is related to the linear instability of the equilibrium configurations and not to the transition to the fundamental state.

As a summary of the result in this Section, Table~\ref{tab:table1} shows the fate of the excited state of several boson stars with different self-interaction coupling and oscillation frequency $\omega$. As shown above, the larger is $\Lambda$, the larger is the space of parameter in which excited states become dynamically robust. Interestingly, there is not a threshold compactness after which excited boson stars are stable, as it is manifest in Figure~\ref{fig13}, but rather a region of the parameter space that grows with increasing $\Lambda$. A different (and approximate) physical criterion for instability will be suggested in Section~\ref{sec4}.

\begin{table}[h!]
\caption{Fate of the excited boson stars (BSs) with different values of the self-interacting coupling $\Lambda$ and oscillation frequency $\omega$.}\label{tab:table1}
\begin{tabular}{c|cccccc}
$\omega$ &0.86 & 0.88 & 0.90 & 0.92 & 0.94 & 0.96\\
\hline
$\Lambda$ = 0 & {\it BH formation}  & {\it BH formation}   & {\it BH formation}   & {\it BH formation}   & {\it BH formation}   & {\it  BH formation}   \\
$\Lambda$ = 25 &  {\it BH formation}  &{\it BH formation}  & {\it BH formation}   & Nodeless BS & Nodeless BS & Nodeless BS  \\
$\Lambda$ = 50 & {\it BH formation}  & {\it BH formation}  & {\it BH formation}  & Nodeless BS & Nodeless BS & Nodeless BS \\
$\Lambda$ = 65 & {\it BH formation}  & {\it BH formation} & {\it BH formation}  & Nodeless BS & Nodeless BS & Nodeless BS\\
$\Lambda$ = 75 & {\it BH formation}  &{\it BH formation}  & {\bf Excited BS} &  {\bf Excited BS} & Nodeless BS & Nodeless BS\\
$\Lambda$ = 85 & {\it BH formation}  &  {\bf Excited BS} &  {\bf Excited BS} &  {\bf Excited BS} &  {\bf Excited BS} & Nodeless BS\\
$\Lambda$ = 100&  {\bf Excited BS} &  {\bf Excited BS} &  {\bf Excited BS} &  {\bf Excited BS} &  {\bf Excited BS} &  {\bf Excited BS}\\
\end{tabular}
\end{table}

\subsection{Dynamical formation}
To complement the stability studies we have just reported, we have also studied the possible dynamical formation of excited (nodeful) boson stars with $\Lambda=100$. Here, the rationale is to start with a spherically symmetric dilute Gaussian distribution of the scalar field \textit{with one node}, of the form:
\begin{equation}
\phi(t=0,r) = \phi_0 \, \left[e^{-(r-r_{0})/\sigma^2} - 0.9\,e^{-(r-r_{0})/(1.2\sigma)^2}\right] \ ,
\end{equation}
where $r_{0}$ is the position of the maximum, $\sigma$ the width, and $\phi_0$ is the amplitude at $t=0$. As illustrative values, we take $r_0 = 0$, $\sigma=60$, and $\phi_0 = 8.5\times10^{-3}$ for $\Lambda=100$ ($\phi_0 = 5.0\times10^{-3}$ for $\Lambda=0$) in the simulations exhibited here.

\begin{figure}[t]
 \includegraphics[width=0.5\textwidth]{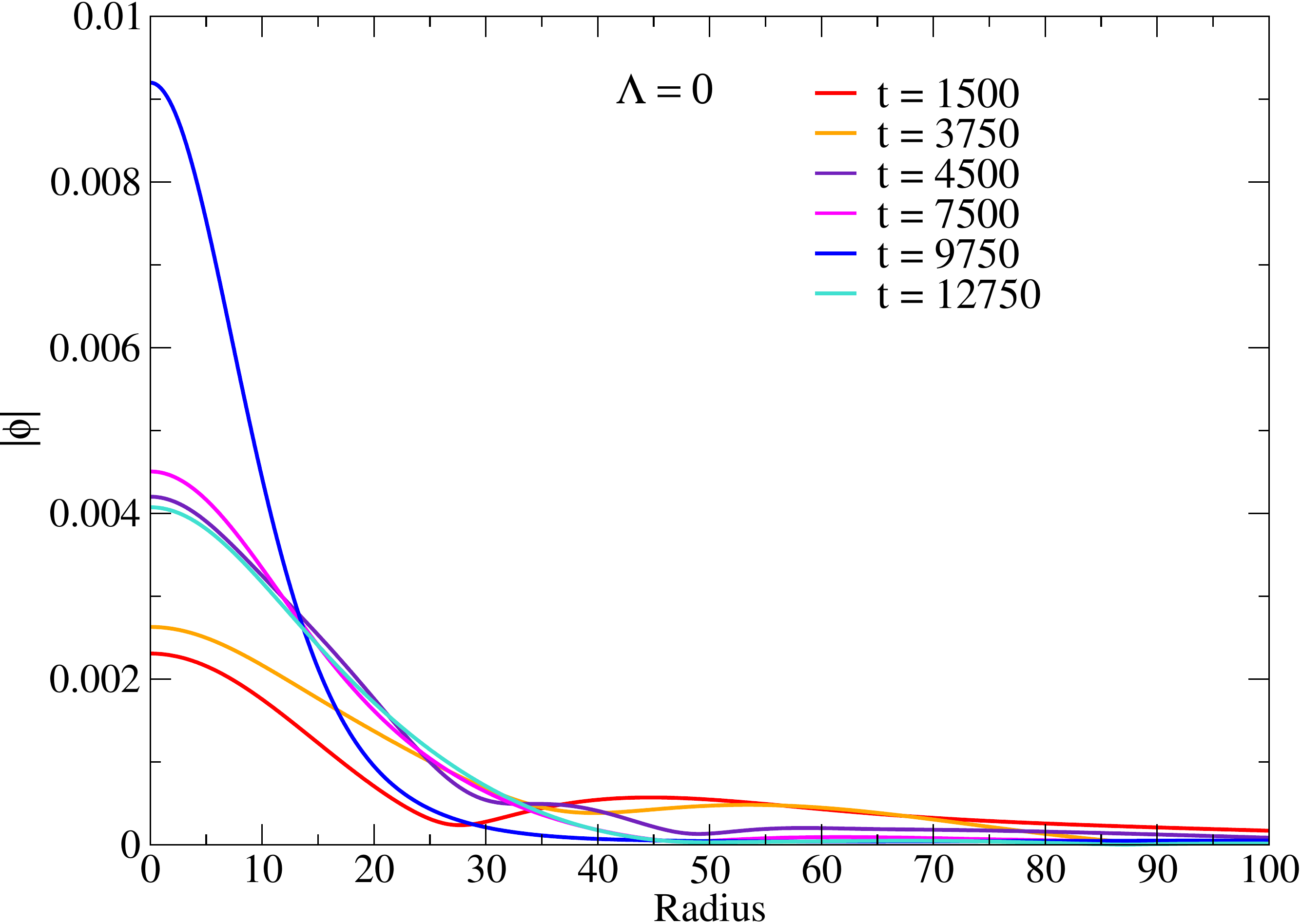}
  \includegraphics[width=0.5\textwidth]{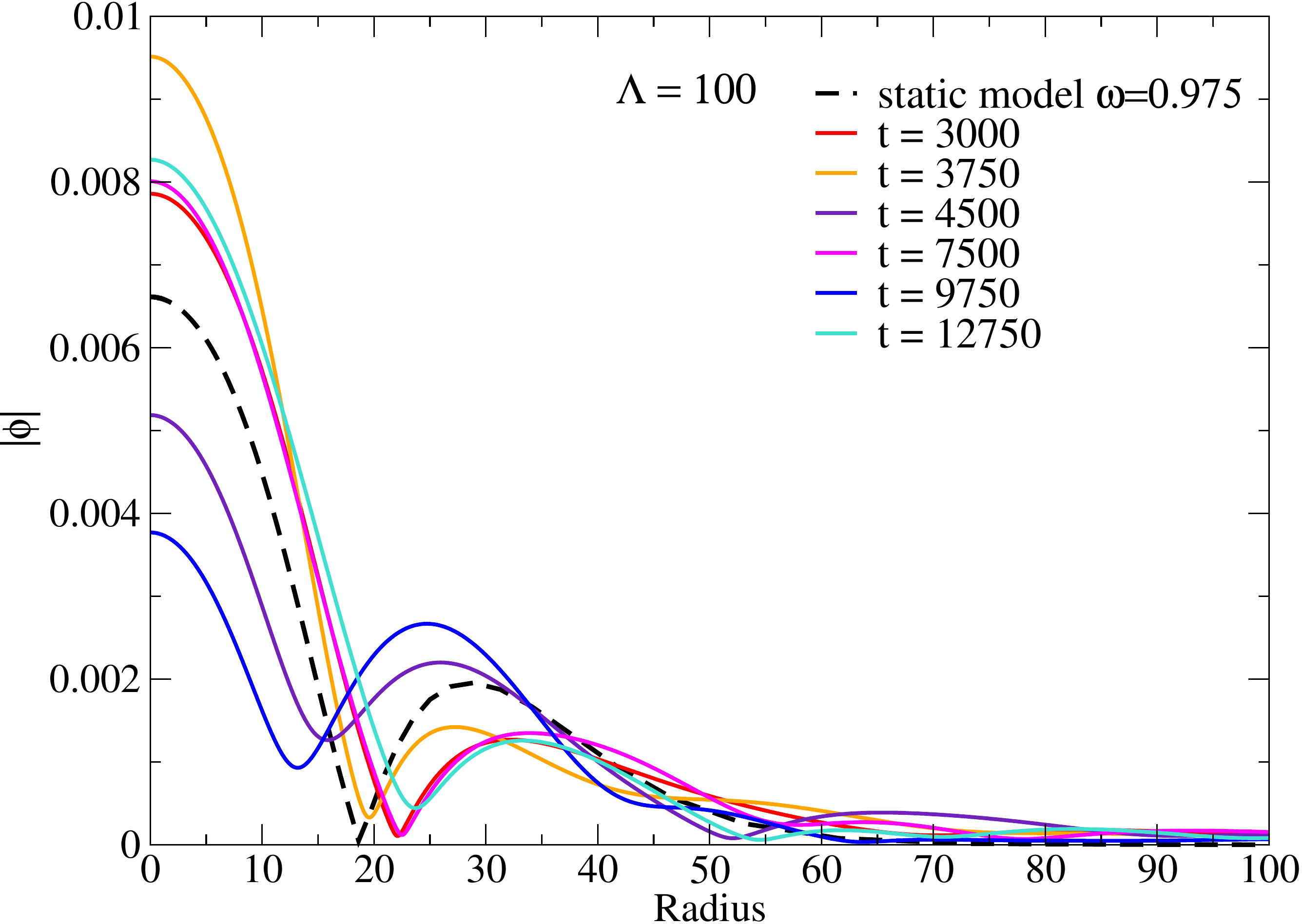}\\
  \includegraphics[width=0.5\textwidth]{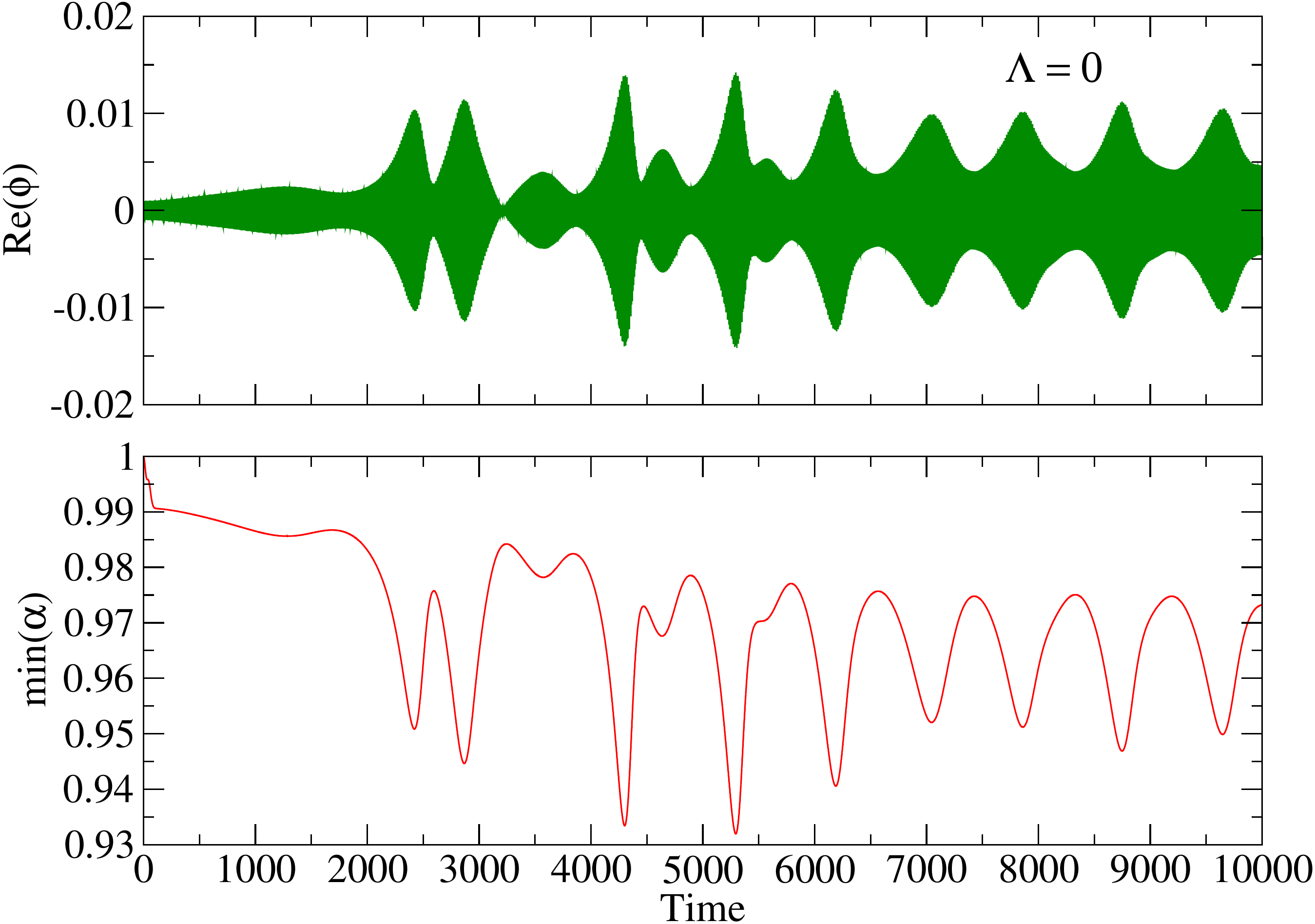}
\includegraphics[width=0.5\textwidth]{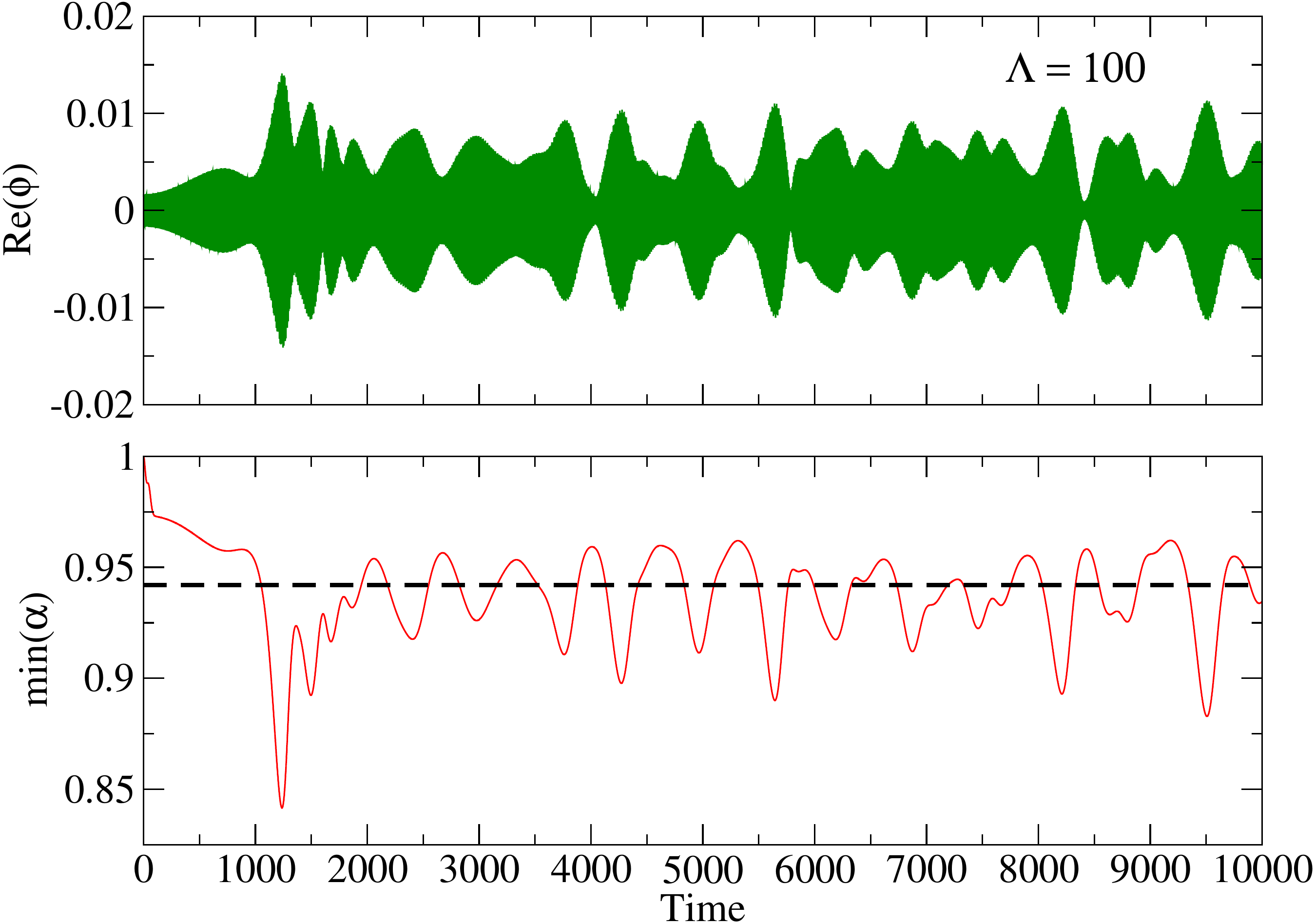}
\caption{Radial profile of the scalar field magnitude $|\phi|$ at different times in the formation scenario with $\Lambda=0$ (top left panel) and $\Lambda=100$ (top right panel). Time evolution of the amplitude of the real part of the scalar field and of the minimum value of the lapse with $\Lambda=0$ (bottom left panel) and $\Lambda=100$ (bottom right panel). The black dashed line gives the minimum value of the lapse of the static model.}
\label{fig:formation}
\end{figure}

We solve the Hamiltonian constraint, Eq.~(\ref{hamiltonian}), as described in previous works~\cite{Sanchis-Gual:2015bh,Sanchis-Gual:2015lje,di2018dynamical,di2020dynamical,DiGiovanni:2021vlu}. The time evolution of the cloud is shown in Figure~\ref{fig:formation} and, for comparison, we show the case with $\Lambda=0$ (left panels) and $\Lambda=100$ (right panels). In both cases, the dilute cloud becomes more compact during the evolutions, approaching  an equilibrium boson star solution, and ejecting the excess energy via gravitational cooling. However, whereas in the $\Lambda=0$ case the initial node of the cloud is lost, and at late times the radial profile of the scalar field is monotonically decreasing, for the $\Lambda=100$ this is not so. One can see that although the star is still vibrating, as it is relaxing towards equilibrium, and thus it does not show the node at all times, it always keeps the second maximum in its relaxation process.  In fact, in the top right panel of Fig.~\ref{fig:formation} we plot a fiducial static model (thick black line) with similar mass ($M=1.092$) and $\omega=0.975$ for comparison, and we observe that the star that is forming from the gravitational cooling mechanism appears to oscillate around the profile of this fiducial model. 

Although it is not possible in the evolutions within this formation scenario to get to very final state, one can clearly appreciate the dependence on $\Lambda$ of the evolutions, and the fact that sufficiently high values of $\Lambda$ keep the shell structure of the energy density, even starting far off from an excited equilibrium solution, whereas for $\Lambda=0$ the scalar field naturally reorganizes itself into a fundamental boson star profile. 

\section{Conclusions}
\label{sec4}
In this work we have investigated the stability of excited boson stars with quartic self-interaction. We found that, as  previously shown in the rotating case, when taking a large self-interaction coupling, the instability of these objects can be quenched up to long computational times. Of course, numerical evolutions as the ones done here cannot \textit{demonstrate} stability. Yet, there the numerical evidence we have presented clearly shows the ``healing" power of self-interactions of the type we have considered. Considering a potential \textit{demonstration} of stability for large self-interactions,  it would be interesting to attempt to tackle it in the large $\Lambda$ limit using approximations such as the ones studied in~\cite{Colpi:1986ye}.

Let us comment that another way to stabilize excited spherical boson stars was studied in the literature~\cite{Bernal:2009zy} (see also~\cite{DiGiovanni:2021vlu}). Here the idea was to study a multi-field model, with one field being left in the fundamental (nodeless) state, and a second field composing an excited (nodeful) state. The resulting spherical boson stars were shown to be stable if the population of the fundamental state was large enough, for a fixed population of the excited state. On the other hand, in the case of multi-field, but not spherically symmetric solutions in~\cite{Sanchis-Gual:2021edp}, it was found that the criterion for stability was associated to  the energy density peak of the composed system becoming shifted towards the origin (see also~\cite{Guzman:2019gqc}). Thus, in both cases stability requires some dominance of the central component, either in terms of Noether charge or energy density. 

\begin{figure}[h!]
\centering
\includegraphics[height=2.27in]{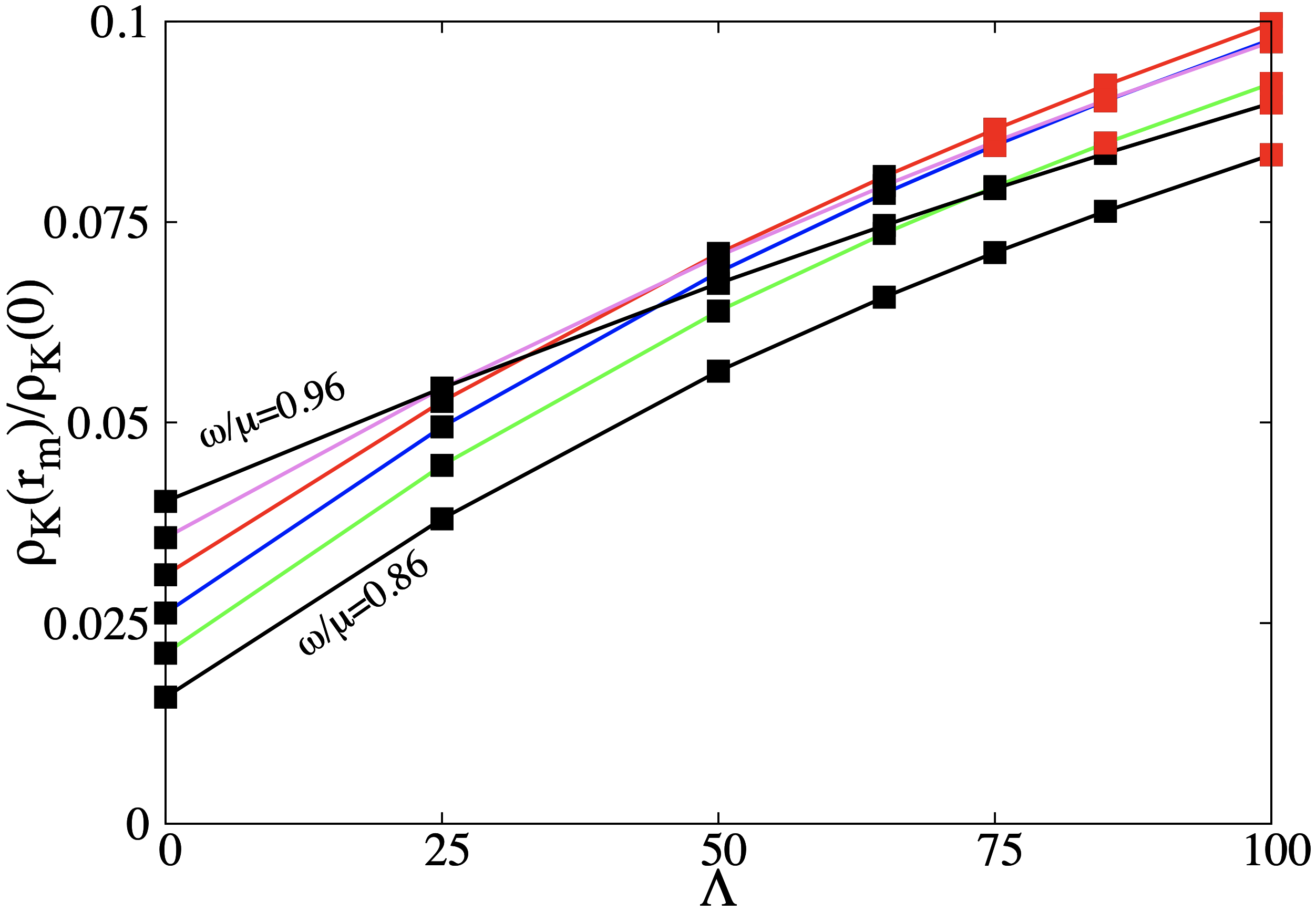}  
\caption{ Ratio of the Komar energy density at the shell radius $r_{\rm m}$ over that at the origin. This ratio increases as $\Lambda$ increases for each value of $\omega$. One can see that stable models only occur above a certain threshold.
}
\label{figratio}
\end{figure}

Attempting to find an interpretation, or at least a physical diagnosis of stability/instability, in the study presented here of spherical excited states with $n=1$, it is convenient to face these stars as a central bosonic ball surrounded by an exterior bosonic shell. Then,  we have observed a suggestive correlation with   the relative energy distributions of the central sphere and the surrounding shell.  This is illustrated in Figure~\ref{figratio}, where we plot the ratio of the Komar energy density~\eqref{ked} at the local maximum which occurs away from the origin (at $r=r_{\rm m}$) over the Komar energy density at the local maximum at the origin.  Figure~\ref{figratio} suggest that stability (roughly) requires the second radial maximum of the energy to be sufficiently \textit{large} as compared to the first one. Interestingly, this seems to follow the opposite trend to the one discussed above for the multi-field stabilization: here there must be a sufficiently \textit{large} energy peak away from the origin for stability.

Bosonic stars have been proposed as dark matter constituents. Moreover, if they do not interact (or interact very weakly) with the electromagnetic field, they could mimic part of the phenomenology of black holes, see $e.g.$~\cite{Herdeiro:2021lwl}, and could also change the properties of fermion stars~\cite{valdez2013dynamical,valdez2020fermion,di2020dynamical,DiGiovanni:2021vlu}. Such objects are still theoretical and have not been observed to date. However, a recent work~\cite{CalderonBustillo:2020srq} found that the head-on collision of vector boson stars (also known as Proca stars) can fit the real gravitational-wave event GW190521, slightly better than the vanilla black hole binary merger.  Studying their non-linear stability is, therefore, important in order to check if they are viable dark objects that could account for part of dark matter, but also may help to understand and make predictions about the stability of similar objects (neutron stars or other exotic compact objects), since interesting parallelisms between neutron stars and boson stars can be drawn, for instance in the development of the bar-mode instability~\cite{baiotti2007accurate,espino2019dynamical,DiGiovanni:2020ror}, despite being described by completely different types of matter.

\section*{Acknowledgements}

We thank Pedro Avelino, Jos\'e Queir\'os, Fabrizio Di Giovanni, and Jos\'e A. Font for useful discussions and valuable comments. This work was supported by the Center for Research and Development in Mathematics and Applications (CIDMA) through the Portuguese Foundation for Science and Technology (FCT - Funda\c c\~ao para a Ci\^encia e a Tecnologia), references UIDB/04106/2020 and UIDP/04106/2020, by national funds (OE), through FCT, I.P., in the scope of the framework contract foreseen in the numbers 4, 5 and 6
of the article 23, of the Decree-Law 57/2016, of August 29,
changed by Law 57/2017, of July 19 and by the projects PTDC/FIS-OUT/28407/2017,  CERN/FIS-PAR/0027/2019 and PTDC/FIS-AST/3041/2020. This work has further been supported by  the  European  Union's  Horizon  2020  research  and  innovation  (RISE) programme H2020-MSCA-RISE-2017 Grant No.~FunFiCO-777740. We would like to acknowledge networking support by the COST Action GWverse CA16104.
Computations have been performed at the Servei d'Inform\`atica de la Universitat
de Val\`encia and the Argus and Blafis cluster at the U. Aveiro.


\bibliographystyle{myutphys}
\bibliography{num-rel2}

\end{document}